\documentclass[table,twocolumn]{aastex62}
\usepackage{amstext}
\usepackage{amsmath}
\usepackage{apjfonts}
\usepackage[capbesideposition={top}]{floatrow}
\usepackage{float}
\usepackage{cancel}
\usepackage{floatrow}
\usepackage{cleveref}
\usepackage{lineno}

\newcommand{\beqar}{\begin{eqnarray}}
\newcommand{\eeqar}{\end{eqnarray}}

\newcommand{\rg}{r_{\rm g}}

\newcommand{\beq}{\begin{equation}}
\newcommand{\eeq}{\end{equation}}

\definecolor{nick}{HTML}{006400}

\newcommand\hammer{\texttt{H-AMR}}
\newcommand\RAPTOR{\texttt{RAPTOR}}
\newcommand\CLOUDY{\texttt{CLOUDY}}
\begin{document}
\title{Changing-Look AGN Powered By Disk Tearing}

\correspondingauthor{Nicholas Kaaz}
\email{nkaaz@princeton.edu}

\author[0000-0002-5375-8232 ]{Nicholas Kaaz}
\affiliation{Princeton Center for Theoretical Sciences, Princeton University, Princeton, NJ 08544, USA}
\affiliation{Princeton Gravity Initiative, Princeton University, Princeton, NJ 08544, USA}

\author[0000-0003-4475-9345]{Matthew Liska}
\affiliation{Center for Relativistic Astrophysics, Georgia Institute of Technology, Howey Physics Bldg, 837 State St NW, Atlanta, GA 30332, USA}

\author[0000-0002-4557-6682]{Charlotte Ward}
\affil{Department of Astronomy \& Astrophysics, 525 Davey Lab, 251 Pollock Road, The Pennsylvania State University, University Park, PA 16802, USA} 

\author[0000-0002-2685-2434]{Jordy Davelaar}

\begin{abstract} 
Changing-look active galactic nuclei (CLAGN) feature order-of-magnitude variability in both the continuum and broad line luminosities on months-to-years long timescales, and are currently unexplained. Simulations have demonstrated that rotating black holes sometimes tear apart tilted accretion disks. These tearing events violently restructure the disk on timescales much shorter than a viscous timescale, hinting at a connection to CLAGN. Here, we show that disk tearing can power changing-look events. We report synthetic observations of an extremely high resolution three-dimensional general-relativistic magnetohydrodynamic simulation of a geometrically thin, tilted accretion disk around a rapidly rotating, $10^8\,M_\odot$ black hole. We perform ray-tracing calculations that follow the thermal disk emission to both a line of sight camera and to a distribution of cameras in a prescribed torus-like broad line region. The continuum photoionizes the broad line region and we calculate the resulting spectrum. Both the continuum and line luminosities undergo order of magnitude swings on months-to-years long timescales. We find shorter, weeks long variability driven by the geometric precession of the inner disk and an intraday quasi-periodic oscillation driven by radial breathing of the inner disk. When the torn disk precesses, it causes asymmetric illumination of the broad line region, driving time-evolving red-to-blue asymmetries of the broad emission lines that may be a smoking gun for disk tearing. We also make predictions for future photometric observations from  \textit{ULTRASAT} and Vera Rubin Observatory given a Shakura-Sunyaev-like effective temperature profile, both of which may play an important role in detecting future changing-look events. Future work should consider nonthermal emission, especially from the time-variable corona.

\end{abstract}


\section{Introduction}
\label{sec:intro}

``Changing-look'' active galactic nuclei (AGN) undergo order of magnitude variations in luminosity on timescales of months to years \citep{shappee_2014,denney_2014,lamassa_2015}. This causes the illumination of the broad-line region (BLR) to become highly time-variable. Although we expect that both the continuum and line emission vary fractionally ($\sim20-40\%$) in most AGN \citep{ulrich_1997,vandenberk_2004}, changing-look AGN are distinguished by the complete (dis)appearance of the broad emission lines. This has upturned the AGN unification paradigm in which the presence or absence of broad lines is determined by our viewing angle \citep{antonucci_1993,netzer_2015}. Instead, the swings in luminosity require either restructuring or obscuration of the innermost accretion flow \citep{ricci_2023}. While obscuration may responsible in some systems, in others it has been ruled out \citep{lamassa_2015,macleod_2016,trakhtenbrot_2019,ricci_2020}; we focus on the latter subset, which we refer to as ``Changing-state'' AGN (CSAGN). The main challenge with CSAGN is that accretion disks usually restructure on a viscous timescale, which is on the order of centuries for canonical thin disks \citep{ss73,pringle_1981}. This is far shorter than the months-to-years long CSAGN timescale \citep{lawrence_2018}, which requires a more unconventional accretion mechanism to explain \citep[e.g., ][]{sniegowska_2020,feng_2021,scepi_2021,mckernan_2022,sierra_2025}. Here, we focus on one possible culprit: accretion disks that are tilted with respect to the BH spin axis. When BHs rotate, they twist up the surrounding space-time. This torques any surrounding gas, causing tilted orbits to begin precessing; in a disk, these torques are communicated by magnetohydrodynamic stresses. This results in a \textit{warp}, meaning that the orientation of the disk varies with radius. Warped disks exhibit significantly different dynamics than planar disks \citep{bardeen_petterson_1975,papaloizou_pringle_1983,pringle_1992,ogilvie_1999}. If the warp is too strong, the disk \textit{tears}  into multiple sub-disks \citep{nixon_2012,liska_2021}. These sub-disks tend to crash into each other and live short, violent lives as they are rapidly consumed. This behavior causes highly variable emission, hinting that torn disks may result in CSAGN \citep[e.g., ][]{raj_2021,Kaaz_2023}. 

Strongly warped accretion disks enter the `bouncing' regime, characterized by extreme scale height variations twice an orbit \citep{fairbairn_2021,kaaz_2025}. At the scale height minima -- which we refer to as ``nozzles'' -- shear layers are compressed and sometimes shocks form, resulting in rapid inflow \citep{Kaaz_2023}. To put this into  context, the inflow rate of canonical thin disks is usually set by the $\alpha$ viscosity parameter, which is theoretically bounded below unity and is usually in the range $\alpha\sim0.01-0.1$ \citep{pringle_1981}. However, if we assume that the orbital energy of the fluid is efficiently dissipated at the nozzles, we find inflow rates strictly higher than $\alpha$ theory predicts. Although the nozzle shock dissipation is distinct from a local viscous stress, it can be useful to write the inflow rate in terms of an ``effective'' viscosity parameter, $\alpha_{\rm eff}$, which is \textit{at least} order unity \citep{kaaz_2025}. The resulting inflow timescale is broadly consistent with CSAGN observations,

\begin{equation}
    \tau_{\rm inflow} \sim 1\,{\rm yr}\,\left(\frac{h}{0.02}\right)^{-2}\left(\frac{\alpha_{\rm eff}}{5}\right)^{-1}\left(\frac{r_{\rm tear}}{10\,\rg}\right)^{3/2}\left(\frac{M_{\rm BH}}{10^8 M_\odot}\right),
\end{equation}

where we have taken characteristic values from \citep[][henceforth `K23']{Kaaz_2023}. In K23, we performed an extremely high resolution general-relativistic magnetohydrodynamic simulation of a tilted accretion disk around a rapidly rotating BH, and studied how tearing and bouncing affect accretion. 

\begin{figure}
    \centering
    \includegraphics[width=\textwidth]{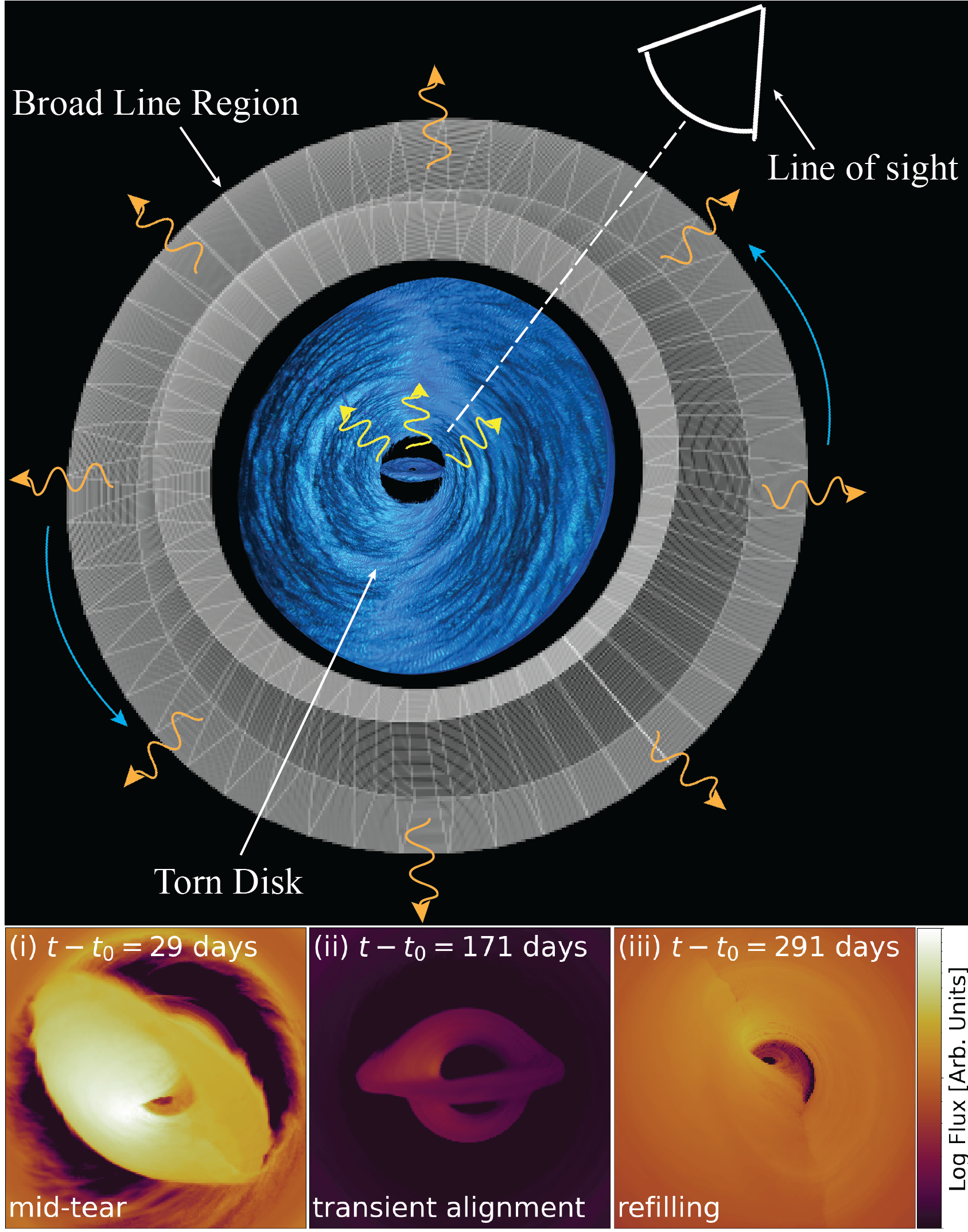}
    \caption{Three-dimensional gas density rendering of a torn disk with a cartoon depiction of the broad line region (BLR). Light colors are high density, dark colors are low density. The majority of the emission emanates from the inner disk (yellow squiggles). This emission photoionizes the BLR, which is idealized as a rotating torus of optically thin gas located $\sim45^\circ$ above the outer disk (white region). Each segment of the BLR then emits lines (orange squiggles) that are Doppler shifted due to the Keplerian motion of the BLR gas. Throughout this paper, we assume a line of sight at a $15^\circ$ inclination. \textbf{Panels (i)-(iii)} Film strip of ray-traced images featuring various phases in the evolution of the inner disk. Panel (ii) is shown at the same time as the 3D rendering in the main panel.}
    \label{fig:3d}
\end{figure}

In this work, we extend K23 by ray-tracing the emission from the warped accretion disk to produce synthetic observations of both the continuum and BLR emission. We show a 3D gas density rendering in Fig.~\ref{fig:3d}, which shows a torn inner disk. We also depict our assumed BLR in white, which is an optically thin torus of rotating gas, located $\sim0.05\,{\rm pc}$ from the BH and extending $\sim45^\circ$ above the accretion disk. The disk photoionizes the BLR gas which then radiates isotropically (orange squiggles). Lines emitted by the BLR are Doppler shifted due to its rotational velocity, which broadens them. We also show a film strip (panels (i)-(iii)) of various phases in the lifetime of a tearing event. Panel (i) shows the disk mid-tear, where the luminosity is high. Panel (ii) shows the inner disk when it has transiently aligned with the BH spin and has not yet been consumed, resulting in a lower luminosity because the warp has died; this image is also shown at the same time as the 3D image in the main panel. Panel (iii) shows the the next tearing cycle where the inner regions are filling in and the luminosity is once again large. 

We structure the paper as follows.
In Section \ref{sec:approach} we describe our methodology, which includes the simulation (\S\ref{sec:approach:sim}), our BLR assumptions (\S\ref{sec:approach:blr}) and our synthetic observations (\S\ref{sec:approach:synth_obs}). In Section \ref{sec:results} we describe our results, including an overview of the tearing-induced variability (\S\ref{sec:results:tearing}), band-dependent features (\S\ref{sec:results:bands}), a quasi-periodic oscillation (\S\ref{sec:results:qpo}) and the response of the broad line region (\S\ref{sec:results:blr}). In Section \ref{sec:disc} we discuss how to observationally determine if a CSAGN is powered by disk tearing (\S\ref{sec:disc:obs}), present caveats to our work (\S\ref{sec:disc:caveats}), and summarize (\S\ref{sec:disc:summary}).


\section{Approach}
\label{sec:approach}

\subsection{H-AMR Simulation}
\label{sec:approach:sim}
We return to the extremely high resolution accretion disk simulation that was previously reported on in K23 and \citet{HAMR}. This accretion disk was simulated by evolving the equations of general-relativistic ideal magnetohydrodynamics (GRMHD) in the code \hammer{} \citep{HAMR}. We report our results after time $t_0\approx 285\,{\rm days}\approx 5\times10^4\,r_{\rm g}/c$. The equations are scale-free, so the simulation can represent any BH mass or density scale. Throughout this work, we take $M=10^8\,M_\odot$ and rescale the gas density (which is initially $\leq 1$) by a factor $\sim10^{-10}\,{\rm g\,cm^{-3}}$. After rescaling, our average mass accretion rate is $1\%$ of the Eddington limit (assuming a 10\% radiative efficiency). 

\textit{Initial conditions.}
We initialize the accretion disk with dimensionless density scale height $H/r=0.02$ around a black hole (BH) with dimensionless spin $a=0.9375$. The vertical profile is a Gaussian in density and locally isothermal, such that the pressure and density scale heights are the same. We use a surface density profile $\Sigma\propto r^{-1}$ that extends from $r=6.5\,\rg$ to $76\,\rg$, where $\rg=GM/c^2$ is the gravitational radius. We initialize the gas with orbital frequency $\Omega=(c/\rg)(r/\rg+a)^{-3/2}$ and insert a purely toroidal field described by a covariant vector potential, $A_\theta\propto(\rho-0.0005)r^2$, where $\rho$ is the fluid-frame gas density. We set the maximum value of $\rho$ to unity and normalize the magnetic field such that the ratio of gas to magnetic pressure is $\beta\approx7$ throughout the disk. The disk is cooled using the  \citet{Noble_Cooling} prescription, which removes the internal energy from a given cell if the gas exceeds a locally isothermal temperature profile $T(r)$. We choose $T(r)$ such that it corresponds to a scale height $H/r=0.02$ and we cool on an orbital timescale. We tilt the entire disk by $65^\circ$ from the equator by rotating it about the $y$ axis. 

\textit{Grid extent and boundary conditions.}
We use a spherical-polar grid that is discretized in ${\rm log}r$, $\theta$ and $\varphi$. The inner radius is $r_{\rm in}=0.9(1+\sqrt{1-a^2})\rg$ and the outer radius is $r_{\rm out}=10^5\rg$. The polar ($\theta$) direction extends from $0$ to $\pi$ and the azimuthal ($\varphi$) direction extends from $0$ to $2\pi$. The inner and outer radial boundaries are outflowing, the $\varphi$ boundaries are periodic, and the $\theta$ boundaries are transmissive. 

\begin{figure*}
    \centering
    \includegraphics[width=\textwidth]{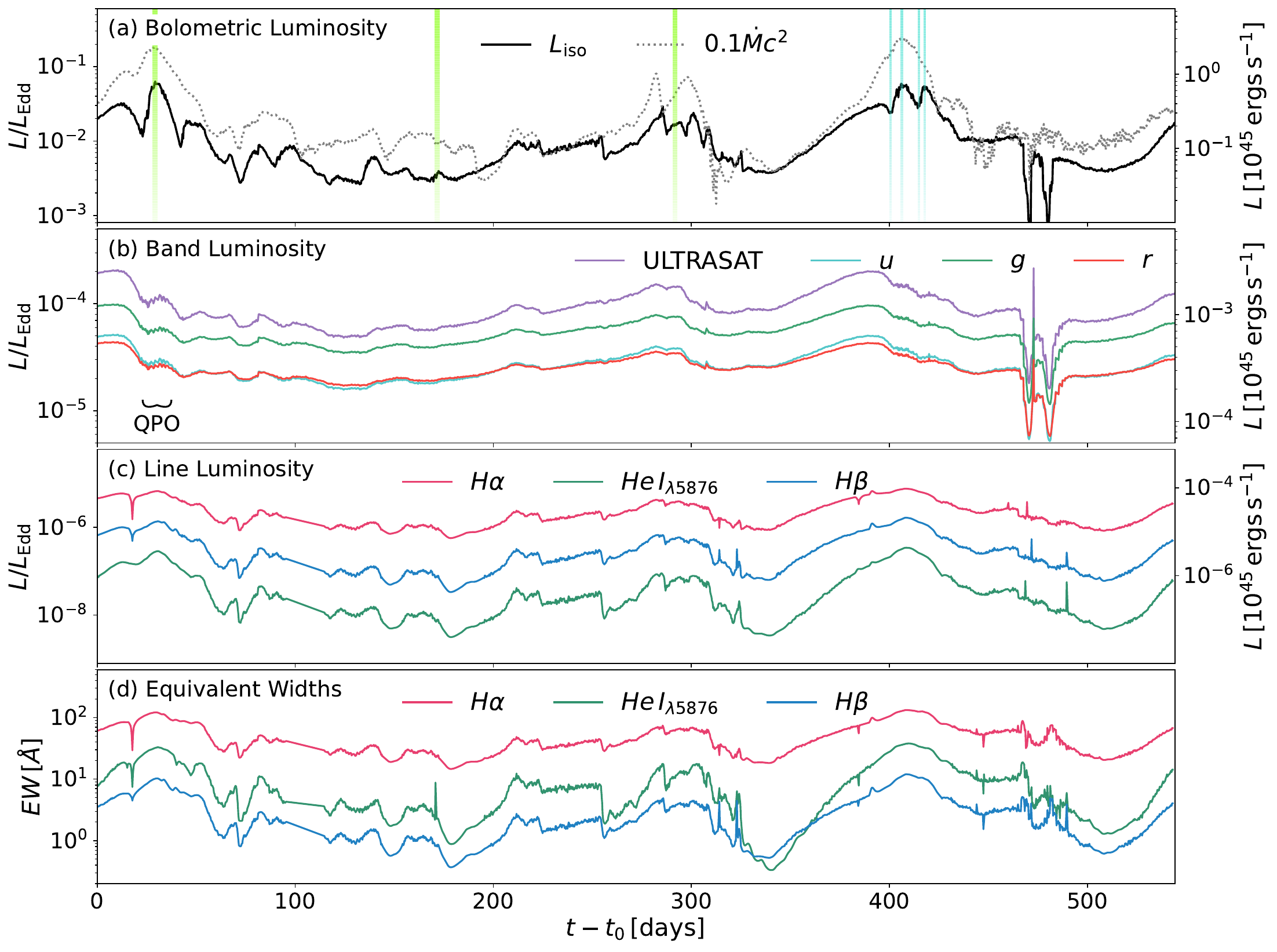}
    \caption{Simulated light curves after time $t_0\approx 285\,{\rm days}\approx 5\times10^4\,r_{\rm g}/c$. Each curve exhibits roughly order-of-magnitude     variability on months-to-years timescales. \textbf{Panel a.} Bolometric isotropic-equivalent luminosity at $15^\circ$ inclination and $0^\circ$ azimuth (black). We also show $\dot{M}$ rescaled to a luminosity assuming 10\% radiative efficiency. $L_{\rm iso}$ mostly tracks $\dot{M}$, but features some weeks-long periodicity due to the geometric precession of the inner disk. One example is highlighted by the vertical blue lines which correspond to the panels in Fig.~\ref{fig:raptor}. Vertical green lines correspond to panels in Figs.~\ref{fig:3d} and \ref{fig:spectra}. \textbf{Panel b.} Same as panel a, except we show the luminosity convolved with filters for \textit{ULTRASAT} \citep{ULTRASAT_2024} and the $u$/$g$/$r$ LSST \citep{LSST_2019} bands. \textbf{Panel c.} Line luminosities in H$\alpha$, H$\beta$ and He I (5876\,\AA). \textbf{Panel d.} Equivalent widths of the lines.}
    \label{fig:lightcurve}
\end{figure*}

\textit{Grid refinement.} We use a base resolution\footnote{\citet{HAMR} and \citet{Kaaz_2023} misreported the base radial resolution in this simulation as $N_r=1728$; $N_r=1680$ is the correct value.} $(N_r\times N_\theta\times N_\varphi)=(1680\times576\times256)$. We then apply two levels of 1D static mesh refinement (SMR) that increase the $\varphi$ resolution to $N_\varphi=1024$ at the equator. These SMR levels \citep[along with block-level ``internal'' derefinement at the poles, see][]{HAMR} makes the cell volumes roughly uniform. On top of the base and 1D SMR layers, we use 3 levels of 3D adaptive mesh refinement (AMR). Small cell sizes coupled with large velocities near the horizon lead to prohibitively small timesteps. To counter-act this, we enable the 1st AMR level at $r\gtrsim2\,\rg$, the 2nd AMR level at $r\gtrsim4\,\rg$, and the 3rd AMR level at $r\gtrsim\,8\rg$. Our AMR criterion locates the disk midplane at each $\varphi$ and ensures that it reaches the highest enabled AMR level. The resulting ``effective'' resolution (i.e., the resolution if the grid was uniformly resolved at the highest level) is $(N_r\times N_\theta\times N_\varphi)=(13440\times4608\times8192)$. This corresponds to about $\approx30$ cells per density scale height. At any given time, we have roughly $\sim14-22$ billion active cells ($\sim2\%$ of the total ``effective'' cell count). In addition to adaptive mesh refinement, we also use 5 levels of local adaptive timestepping, which allows different blocks (even at the same AMR
level) to evolve at different timesteps \citep{LocalAdaptiveTimestepping}.

\subsection{Broad Line Region Assumptions}
\label{sec:approach:blr}

We do not simulate the BLR, which exists far from the BH. Instead, we prescribe the BLR given a few simplifying assumptions that are broadly consistent with observed BLR properties \citep{peterson_2006,cackett_2021}. The BLR exists near a radius, 
\begin{equation}
    R_{\rm BLR} = 10^4\,\rg\approx 0.05\,{\rm pc}\left(\frac{M}{10^8\,M_\odot}\right),
\end{equation}
and at inclination $I_{\rm BLR}=45^\circ$. We take the equator to be the midplane of the outer disk ($60\,\rg\lesssim r\lesssim80\,\rg$). The BLR is axisymmetric and consists of optically thin gas orbiting at the Keplerian velocity,
\begin{equation}
    v_{\rm BLR}= 3000\,{\rm km\,s^{-1}}\left(\frac{R_{\rm BLR}}{10^4\,\rg}\right)^{-1/2}
\end{equation}
The choice of a torus-like BLR geometry is consistent with optical interferometry of quasars \citep{GRAVITY_2018}, but we note that in other AGN a wind-like geometry may be more appropriate \citep[e.g., ][]{murray_1995}. We assume the BLR has a constant Hydrogen number ($n_{\rm BLR}=10^{11}\,{\rm cm}^{-3}$) and column ($N_{\rm BLR}=10^{22}\,{\rm cm}^{-2}$) density. The BLR has a slab-like geometry, where the slab thickness in the radial direction is $\Delta s \approx N_{\rm BLR}/n_{\rm BLR}=10^{11}\,{\rm cm}$. The BLR has polar extent $\delta \theta = \pm 5^\circ$ about its inclination resulting in a $\approx12\%$ covering factor. The BLR gas has solar metallicity. The only source of heat in the BLR is irradiation from the accretion disk. We implicitly assume that the recombination timescale of the BLR is short compared to variations in the irradiating flux. 

 We take our line of sight relative to the BH to be oriented along the vector,
\begin{equation}
    \hat{n}_{\rm obs} = \left(\sin I_{\rm obs} \cos\varphi_{\rm obs},\,\sin I_{\rm obs}\sin\varphi_{\rm obs},\,\cos I_{\rm obs} \right),
\end{equation}

where we take $\varphi_{\rm obs}=0^\circ$ and $I_{\rm obs}=15^\circ$ throughout.
Similarly, the BLR gas relative to the BH moves in the direction, 
\begin{equation}
    \hat{v}_{\rm BLR} = \left(\cos I_{\rm BLR} \cos\varphi_{\rm BLR},\,\cos I_{\rm BLR}\sin\varphi_{\rm BLR},\,\sin I_{\rm BLR} \right),
\end{equation}
where we take $I_{\rm BLR}=45^\circ$ throughout and vary $\varphi_{\rm BLR}$. The relative velocity between our line of sight and the BLR gas is then,
\begin{equation}
    v_{\rm rel} = v_{\rm BLR}\hat{n}_{\rm obs}\cdot\hat{v}_{\rm BLR}
\end{equation}
such that the spectra emitted by the gas is Doppler shifted by a factor $\Delta \nu = (v_{\rm rel}/c)\nu_0$ where $\nu_0$ is the frequency of the light emitted in the rest-frame of the BLR gas. We neglect the thermal and turbulent motions of the BLR gas.

\subsection{Synthetic Observations}
\label{sec:approach:synth_obs}

\textit{Ray-tracing.} We use the radiative transfer software \RAPTOR{} \citep{RAPTOR} to ray-trace our \hammer{} simulation. Since our simulation is not explicitly radiative, we must make assumptions about the thermal properties of the disk. We describe these assumptions in Appendix \ref{app:temp} and review the salient details here, 
\begin{itemize}
    \item We assume that the disk is dominated by a scattering opacity and emits thermally. 
    \item We prescribe an effective temperature\footnote{In practice, simply setting the radiation and gas temperatures equal to each other will result in luminosities that are not self-consistent; especially in shocks, the gas temperature may greatly exceed the radiation temperature.} such that the radiative flux leaving the disk is consistent with our cooling function. 
    \item We exponentially damp the effective temperature in optically thin regions. We do this because these regions depart from our blackbody assumption and we cannot reliably prescribe their thermodynamics. 
    \item We ray-trace our images using backwards integration in the fast light approximation. This means that we start with our rays in the image plane and follow them backwards on null geodesics until they reach the photosphere, and that each image is constructed assuming the simulation is static while the rays propagate.
    \item  Our snapshots are separated by a time interval $50\,r_{\rm g}/c\sim7\,{\rm hrs}$. We choose an image plane width $80\,r_{\rm g}\times80\,r_{\rm g}$ covered by $400\times400$ pixels. The line of sight camera is placed at $\varphi_{\rm obs}=0^\circ$ and $I_{\rm obs}=15^\circ$. We uniformly distribute BLR cameras at every $30^\circ$ in azimuth ($\varphi_{\rm BLR}$) at a fixed inclination $I_{\rm BLR}=45^\circ$ and perform a ray-tracing calculation for each. We interpolate our results at intermediate values of $\varphi_{\rm BLR}$ when computing broadened spectra. 
    \item We then use version 23 of \CLOUDY{} \citep{CLOUDY23} to calculate emission from the BLR. We start with the \RAPTOR{} calculations at each camera in the BLR, which provides us with an incident flux from the AGN, which we assume is the only source of heat in the BLR. \CLOUDY{} calculates the thermodynamic state of each cloud in photoionization equilibrium, providing us with an emitted spectrum. We assume each cloud emits isotropically and we neglect reflected light. 
\end{itemize}


\section{Results}
\label{sec:results}
\subsection{Tearing Cycles}
\label{sec:results:tearing}
\begin{figure}
    \centering
    \includegraphics[width=\textwidth]{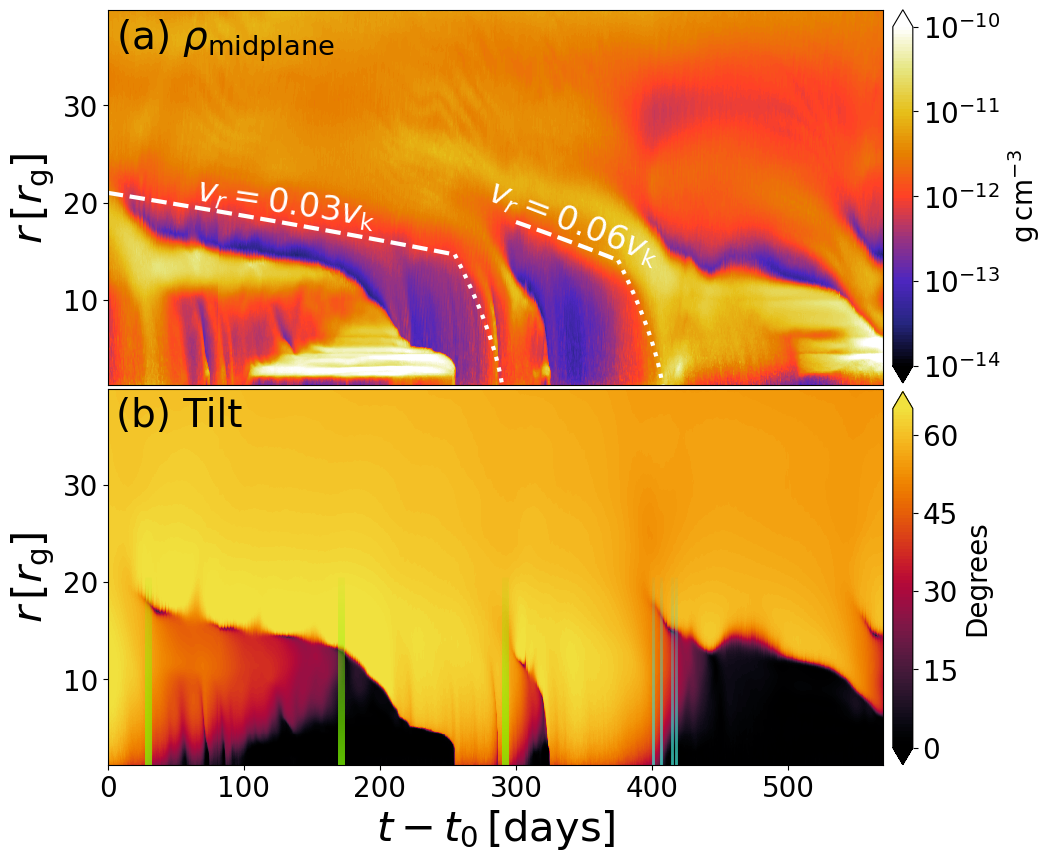}
    \caption{Spacetime diagram of disk midplane density and tilt, which reveals repeated tearing cycles and, sometimes, transient alignment of the inner disk. \textbf{Panel a.} Midplane density, $\rho_{\rm midplane}$. Low-density (purple) regions indicate gaps between sub-disks or a recently-consumed inner disk. We have highlighted two tears with white lines; the outer disk will refill the inner region at a rate $v_r\gtrsim10^{-2}v_{\rm k}$ (white dashed lines) and eventually lead to plunging flows with $v_{\rm r}\gtrsim 10^{-1}v_{\rm k}$ (white dotted lines), which are coincident peaks in the luminosity (Fig.~\ref{fig:lightcurve}). \textbf{Panel b.} Tilt angle, $T$. Usually, $T\sim60-65^\circ$, but the inner regions sometimes transiently align ($\sim100-200$ days and $\sim450-550$ days). The transiently-aligned phases are coincident with the troughs in the light curves (Fig.~\ref{fig:lightcurve}. Green vertical lines correspond to panels in Fig.~\ref{fig:3d} and Fig.~\ref{fig:spacetime}; blue vertical lines correspond to panels in Fig.~\ref{fig:raptor}.}
    \label{fig:spacetime}
\end{figure}

Figure \ref{fig:lightcurve} shows various light curves: the bolometric luminosity, the rescaled mass accretion rate, the luminosity in certain photometric bands, selected broad line luminosities and equivalent widths. We show the isotropic-equivalent bolometric luminosity, $L_{\rm iso}$, in Fig.~\ref{fig:lightcurve}a. This curve features a handful of peaks and troughs over $\sim600$ days. The properties of the tearing cycles change from event to event; it is useful to compare with the spacetime diagram of midplane density, $\rho_{\rm midplane}$, depicted in Fig.~\ref{fig:spacetime}a. Here, low-density (purple) regions indicate either a tear between sub-disks or a near-horizon cavity that exists between tearing cycles. Tearing generally occurs at radii $\lesssim20\,r_{\rm g}$. The first tearing cycle is long-lived, lasting for $\sim250$ days.  At $\sim {\rm 30}$ days, the inner disk is mid-tear, and $L_{\rm iso}$ peaks (marked by the first green vertical line in Fig.~\ref{fig:lightcurve}a; see also Fig.~\ref{fig:3d}i). The sub-disk exists for long enough to transiently align with the BH spin from about $\sim100-250$ days, as evidenced by the small tilt angle in Fig.~\ref{fig:spacetime}b. When the disk aligns, it is no longer warped, and accretion is less violent. This phase is shown in 3D in Fig.~\ref{fig:3d}, in a ray-traced image in Fig.~\ref{fig:3d}ii, and is marked by the second green vertical line in Fig.~\ref{fig:lightcurve}a. When the disk aligns, its luminosity decreases. Fig.~\ref{fig:spacetime}a also shows the outer disk filling in at a rate $v_{\rm r}\sim0.03v_{\rm k}$ (first white dashed line), and once the previous tearing cycle completes, a new tear begins (first white dotted line), leading to dynamical inflow with $v_{\rm r}\sim0.3 v_{\rm K}$ (the processes that drive the rapid inflow are studied in K23). This tear is small (Fig.~\ref{fig:3d}iii) leading to a smaller peak luminosity (third green vertical line in Fig.~\ref{fig:lightcurve}a) and a shorter overall tearing cycle. 

The differences from tearing cycle to tearing cycle have implications for the observational classification of any variability driven by tearing. For instance, quasi-periodic eruptions -- another form a violent AGN variability -- feature periodicities that are highly coherent \citep[e.g., ][]{miniutti_2019} in comparison to the light curves reported here. Our results suggest that tearing powers extreme variability that is much less coherent \citep[although,][do find coherent periodicities using smoothed-particle viscous hydrodynamic simulations]{raj_2021}. Still, there are specific features that distinguish tearing from other types of variability. For instance, compare $L_{\rm iso}$ to $\dot{M}$ in Fig.~\ref{fig:lightcurve}a. The two curves largely track each other because the luminosity is powered by accretion. However, if we turn to the peaks at $\sim30$ days and $400$ days, we see extra variability in $L_{\rm iso}$ but not in $\dot{M}$. This is because the inner sub-disk is precessing in and out of our line of sight at these times. We can see this in the ray-traced images in Fig.~\ref{fig:raptor} (with their times marked by the four vertical blue lines in Fig.~\ref{fig:lightcurve}a), where the inner sub-disk precesses by about $180^\circ$ (and decreases in tilt from about $50^\circ$ to $30^\circ$, see Fig.~\ref{fig:spacetime}b). Since the inner sub-disk tends to be short lived, tearing cycles may feature only a few precession periods. The precession-induced variability superimposed on the longer-term variability may be a detectable signature of disk tearing.

\begin{figure}
    \centering
    \includegraphics[width=0.9\textwidth]{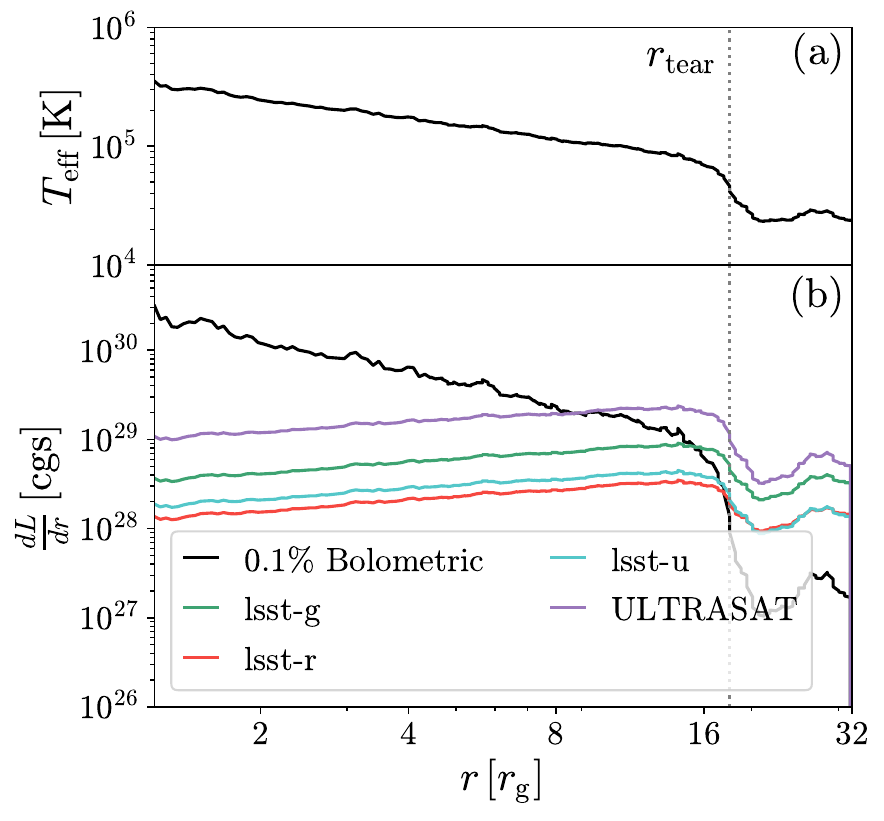}
    \caption{Effective temperature and luminosity per unit radius at time $t-t_0=40\,{\rm days}$. \textbf{Panel a.} The effective temperature of the disk as a function of radius. We also have labeled the radius of the tear, $r_{\rm tear}$. \textbf{Panel b.} The luminosity per unit radius, $\frac{dL}{dr}$ (Eq.~\ref{eq:dLdr}), as a function of radius. We show curves both for the band-dependent luminosity and ($0.1\%$ of) the bolometric luminosity, where we have assumed that each radius of the disk radiates as a blackbody. Note that theses curves are calculated from the data, not ray-traced, so that there are no viewing angle effects. }
    \label{fig:teff}
\end{figure}

\subsection{Band-Dependent Features}
\label{sec:results:bands}

In Fig.~\ref{fig:lightcurve}b, we show light curves convolved with filters for \textit{ULTRASAT} \citep{ULTRASAT_2024} and the Vera Rubin Observatory's $u$, $g$ and $r$ bands \citep[]{LSST_2019}. which peak near $2550$, $3800$, $5370$ and $6770\,$\AA, respectively. We see that the luminosity is usually stronger and more variable at shorter wavelengths (however, here the \textit{u} band emission is weaker than the \textit{g} band because of its lower overall sensitivity). This dependence is expected, since shorter wavelengths track smaller radii where the disk is more luminous and evolves more  quickly. The bolometric luminosity is still more variable than even the \textit{ULTRASAT} light curve since our selected bands do not reach the peak temperature of the emission. This is made clear in Figure \ref{fig:teff}, which is depicted at time $t-t_0=40\,{\rm days}$. In Fig.~\ref{fig:teff}a, we show the effective temperature of the disk as a function of radius, which achieves a peak temperature of $\sim3\times10^5\,{\rm Kelvin}$ and drops off beyond the tearing radius $r_{\rm tear}\approx16.5\,\rg$. In Fig.~\ref{fig:teff}b, we plot the luminosity emitted per unit radius,
\begin{equation}
    \frac{dL}{dr} = 2\pi r \int T(\lambda) B_\lambda(T_{\rm eff}(r)) d\lambda,
    \label{eq:dLdr}
\end{equation}
where $T(\lambda)$ is a stand-in function for a given filter. We show $\frac{dL}{dr}$ for each band and one with the rescaled bolometric luminosity (where we set $T=0.001$). We can see that while the bolometric luminosity peaks at the inner edge of the disk, each of the band peak closer to the outer edge of the torn inner disk. 

We emphasize that our temperature prescription is necessarily ad hoc, and fully radiative simulations are required to accurately estimate the effective temperature. Furthermore, it is well-known that the temperatures of AGN in general are lower than theoretically predicted \citep{lawrence_2012}, peaking near $\sim3\times10^4\,{\rm K}$ \citep{zheng_1997} -- a factor of ten lower than our calculations suggest. However, while theoretical models predict spectral energy distributions (SEDs) that provide enough ionizing flux to power the BLR, observed AGN SEDs do not \citep{netzer_1985}. It is possible, then, that the BLR sees the theoretical AGN SEDs rather than the ones we observe. The truth here is murky, so any band-dependent information that we predict must be interpreted with caution. 

\subsection{An Intra-day Quasi-Periodic Oscillation}
\label{sec:results:qpo}

\begin{figure}
    \centering
    \includegraphics[width=\textwidth]{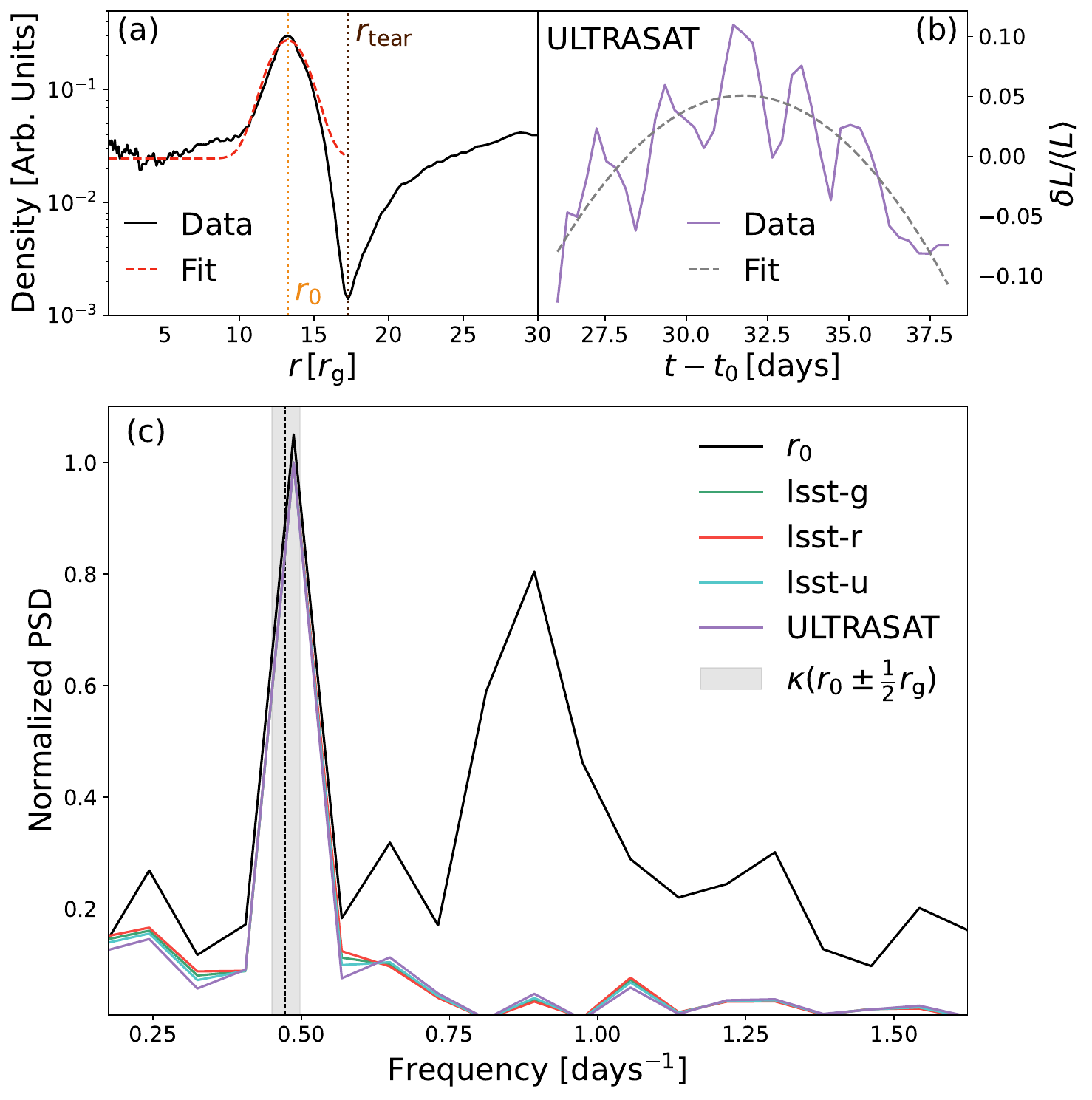}
    \caption{We show a quasi-periodic oscillation and complementary plots that occur at $26\lesssim t-t_0\lesssim40\,{\rm days}$ . \textbf{Panel a.} We show the midplane density profile at time $t-t_0=40\,{\rm days}$. We have fit the density of the inner disk ($r<r_{\rm tear}$) to a Gaussian profile (red). We have marked both the radius of the tear, $r_{\rm tear}$, and the centroid of the Gaussian, $r_0$, with vertical lines. \textbf{Panel b.} We show the fractional variation in the luminosity, $\delta L/\langle L\rangle$, along with a quadratic fit to it during the QPO (see also the QPO annotation in Fig.~\ref{fig:lightcurve}b). Our PSD is calculated by subtracting the luminosity variation from this fit. \textbf{Panel c.} The (normalized) PSD during the QPO for each band, depicting a clear peak at $\sim0.48\,{\rm days}^{-1}$. We also show the PSD of $r_0$, which exhibits a peak at the same frequency as the flux. The shaded region shows the radial epicyclic frequency, $\kappa$, within $\pm\frac{1}{2}r_{\rm g}$ of $r_0$, which is near the QPO frequency. This supports our interpretation that the QPO is powered by axisymmetric oscillations of the inner disk that are driven by the tear \citep[see also][]{gibwa_2023}.}
    \label{fig:qpo}
\end{figure}

There is an interesting feature in the bands shown in Fig.~\ref{fig:lightcurve}b: at roughly $\sim30-40\,$ days, the luminosity fluctuates coherently at the level of a few percent. This is a robust quasi-periodic oscillation (QPO) arising from axisymmetric radial oscillations of the inner, torn sub-disk \citep[e.g., ][]{blaes_2006}. These oscillations were originally identified in this simulation by \citet{gibwa_2023}. In a recent work, this simulation was also ray-traced assuming a stellar-mass BH using a separate methodology by West et al 2025 (upcoming), who discovered that these oscillations can give rise to high-frequency QPOs in X-ray binaries (XRBs). We present our analysis of this feature in Figure ~\ref{fig:qpo}. In Fig.~\ref{fig:qpo}a, we show the midplane density profile at $t-t_0=40\,{\rm days}$ (black), where we can see that the inner disk is concentrated at a radius $r_0\approx13\,r_{\rm g}$ and the tear is at $r_{\rm tear}\approx 17\,r_{\rm g}$. We fit the density profile of the inner disk to a Gaussian profile centered at $r_0$ (red),
\begin{equation}
\rho_{\rm midplane}^{\rm (fit)}(r,t)|_{r<r_{\rm tear}} = \rho_0 + \rho_1\exp\left(-\frac{(r-r_0)^2}{2\sigma^2}\right)
\end{equation}
In Fig.~\ref{fig:qpo}b, we show the fluctuations of the ULTRASAT light curve during the QPO (purple) along with a quadratic fit (gray). The fluctuations about the fit are on the order of a few percent. We subtract the fit from the data to obtain the oscillating signal and plot the resulting power spectral density (PSD) in Fig.~\ref{fig:qpo}c. Each band shows a prominent signal at $\sim0.48\,{\rm days}^{-1}$. In black, we plot the PSD of the midpoint, $r_0(t)$, of the inner sub-disk. There is a strong peak at exactly the same frequency as the QPO. Additionally, in gray, we show that the (lab-frame) radial epicyclic frequency near $r_0$, $\kappa(r_0\pm\frac{1}{2}r_{\rm g})$, is consistent with our signal. This suggests that indeed, this QPO is due to radial oscillations of the torn inner disk, consistent with \citet{gibwa_2023} \citep[see also][, who suggesting that these oscillations could be responsible for quasi-periodic eruptions]{raj_2021}. We also note that $r_0$ features a second peak at a somewhat higher frequency, which we do not investigate in this work.

We speculate that the oscillations of the inner disk are forced by time-dependent interactions with the outer disk. However, it remains unclear how exactly this plays out beyond a qualitative level. Although we observe multiple tearing events in the simulation, the QPO during this particular tearing event is by far the strongest. This suggests that the strength of the QPO depends sensitively on the configuration of the torn disk, which varies from event to event (e.g., Fig.~\ref{fig:spacetime}).

Intraday cadence observations of AGN are not common, so this signal can be easily missed. We also do not know what the band dependence of the signal is in a real disk. Here, the signal is strongest at temperatures probing the outer edge of the inner torn disk ($\sim10-20\,r_{\rm g}$ in this simulation). As discussed in Sec.~\ref{sec:results:bands}, the peak effective temperature of our disk, although consistent with canonical thin disks \citep[][]{ss73}, is a factor of ten higher than the peak temperature of observed AGN SEDs. This makes the band-dependence of the QPO unclear. Still, given the strength of the signal, high cadence observations of variable AGN -- especially CSAGN -- may reveal such a QPO.

\begin{figure}
    \centering
    \includegraphics[width=\textwidth]{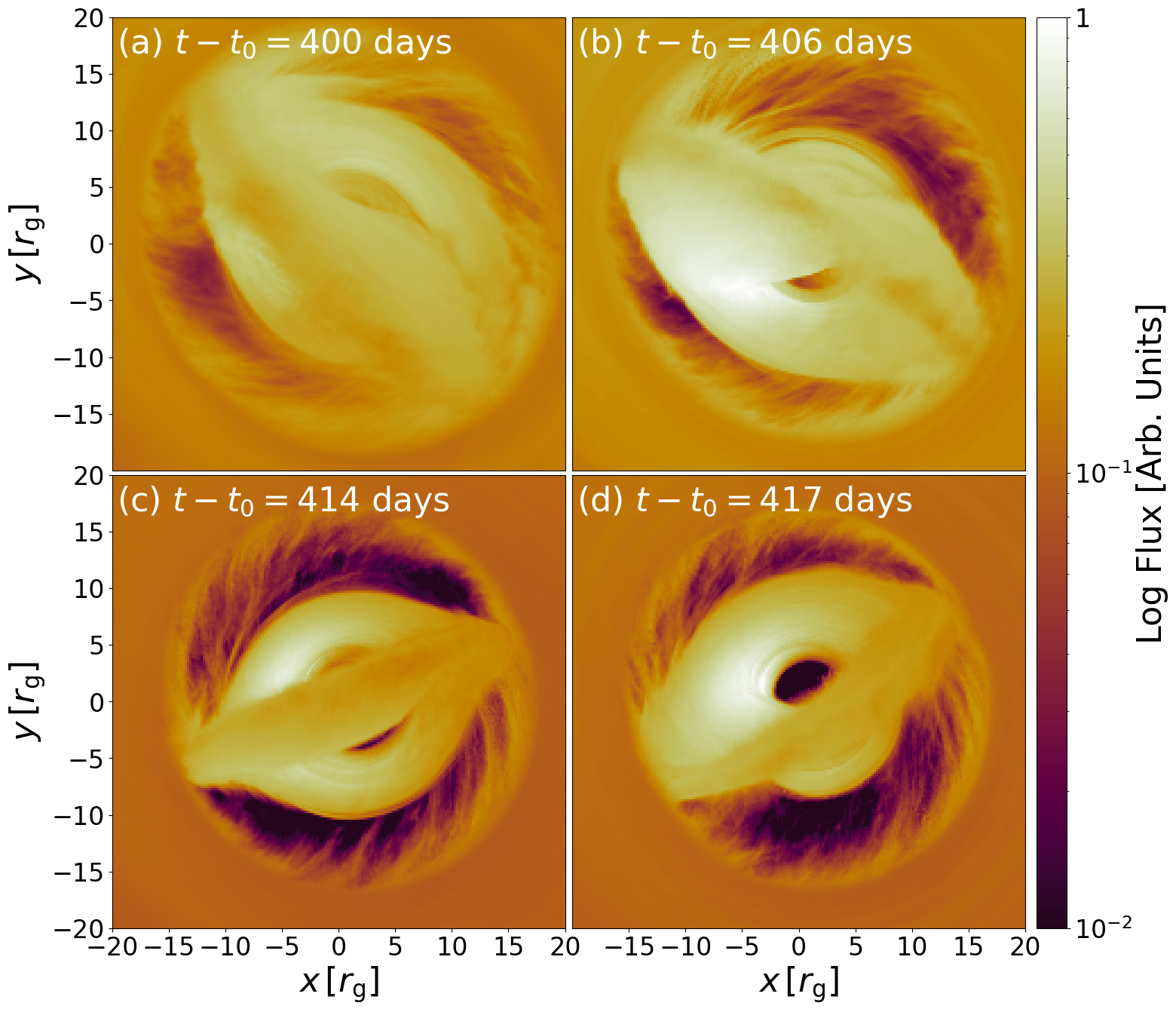}
    \caption{Ray-traced images of the precessing inner disk (corresponds to vertical green lines in Fig.~\ref{fig:lightcurve}a and Fig.~\ref{fig:spacetime}b). We can see a $\sim180^\circ$ geometric precession from panel (a) to (d), which super-imposes weeks-long periodicity in the light curve on top of the months-to-years long  variability.}
    \label{fig:raptor}
\end{figure}

\subsection{Response of the Broad-Line Region}
\label{sec:results:blr}

\begin{figure}
    \centering
    \includegraphics[width=\textwidth]{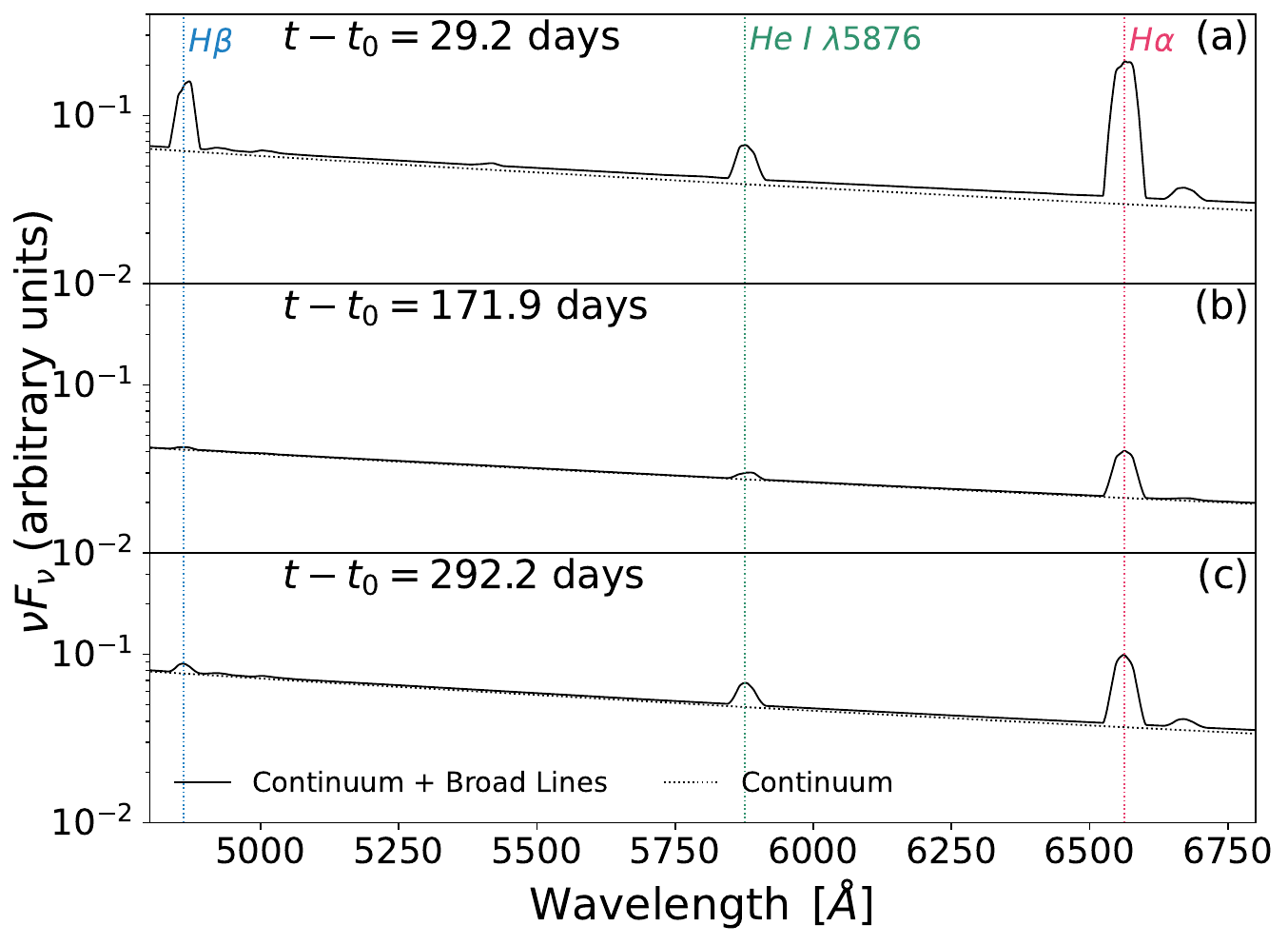}
    \caption{Time series of synthetic optical spectra. We clearly see a temporary (near) disappearance of the broad H$\alpha$, H$\beta$ and He I ($\lambda=5876\,$\AA) lines within a year, which defines this system as a CLAGN. The panels correspond to the green curves in Figs.~\ref{fig:lightcurve} and Fig.~\ref{fig:spacetime} and the ray-traced images in Fig.~\ref{fig:3d}. Panels (a) and (c) are near maxima in the light curves while panel (b) is near a minimum.}
    \label{fig:spectra}
\end{figure}

As described in Sec.~\ref{sec:approach:synth_obs}, in addition to our line of sight camera we also place cameras at regular azimuths within the BLR (the BLR geometry is sketched in white in Fig.~\ref{fig:3d}). After ray-tracing the emitted flux to each camera, we perform \CLOUDY{} calculations to determine the resulting spectrum. We then integrate the contribution to the emission lines from each cloud, interpolate in azimuth, and apply Doppler shifts based on the Keplerian motion of the BLR to compute the composite BLR spectrum. In Fig.~\ref{fig:lightcurve}c, we plot the resulting H$\alpha$, H$\beta$ and He I line luminosities as a function of time. Each line exhibits roughly order of magnitude variability over the span of nearly two years, consistent with observed CSAGN. Fig.~\ref{fig:lightcurve}d shows the equivalent widths of the lines, which measure how strong the broad lines are relative to the continuum. The behavior of the equivalent widths is nearly the same as the line luminosities. This is because the optical luminosity changes less dramatically than the ionizing flux, as evidenced by the muted variability in the optical bands shown in Fig.~\ref{fig:lightcurve}b. In Fig.~\ref{fig:spectra}, we provide a time series of the optical spectra. Each panel corresponds respectively to Fig.~\ref{fig:3d}i-iii and is marked by the vertical green lines in Figs.~\ref{fig:lightcurve}a and ~\ref{fig:spacetime}b. We clearly see that each of the labeled lines (H$\alpha$, H$\beta$ and He I) begin strong, dip to become nearly invisible, and then turn back on. This order-of-magnitude broad line variability defines the system as a CSAGN.

\begin{figure*}
    \centering
    \includegraphics[width=\textwidth]{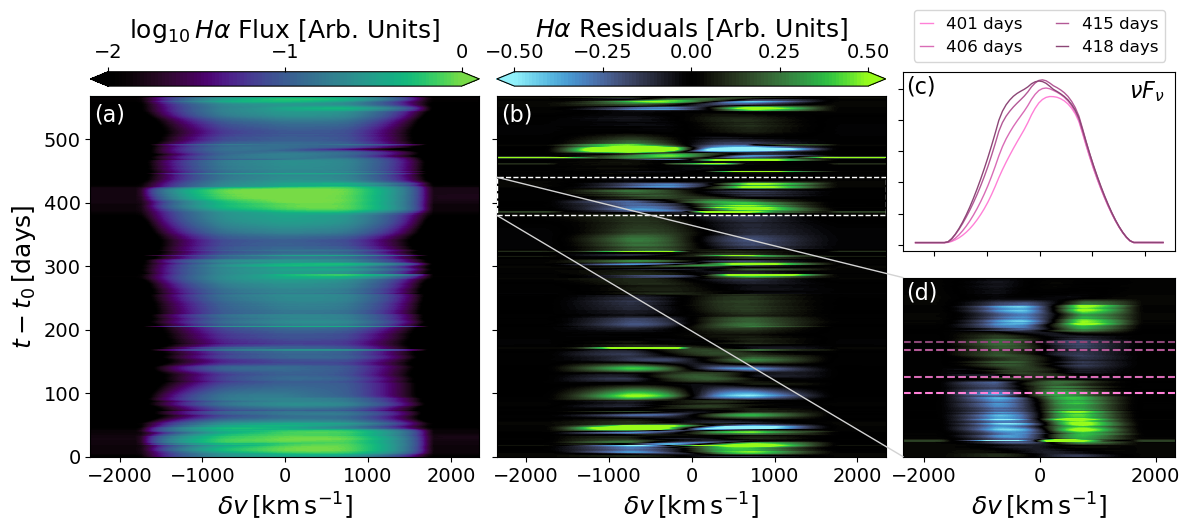}
    \caption{Evolving asymmetries in the H$\alpha$ broad line. \textbf{Panel a.} We depict the profile of the H$\alpha$ broad line as a function of time (see Fig.~\ref{fig:lightcurve}c for the integrated profile as a function of time). \textbf{Panel b.} We show the residuals of the profile shown in panel a, revealing a rich evolution of the broad line asymmetries that trace the evolution of the inner disk (see Sec.~\ref{sec:results:blr} for how we calculate the residuals). \textbf{Panel c.} We show the H$\alpha$ profile at the times coincident with the tearing event depicted in Fig.~\ref{fig:raptor}, wherein the precession of the inner disk drives a shifting asymmetry. \textbf{Panel d.} we zoom-in on the residuals plot to highlight the tearing event; the shaded pink lines correspond to those shown in panel c.}
    \label{fig:halpha_spacetime}
\end{figure*}

We also note that the weeks-long precession induced variability (see Fig.~\ref{fig:raptor} or the blue vertical lines in Fig.~\ref{fig:3d}) is absent from the line luminosities. This is because the BLR covers all azimuths, while the line of sight is at a fixed azimuth. Geometric precession makes the flux strongly dependent on azimuth, causing variability in the line of sight luminosity but not the BLR luminosity. This is interesting, as it suggests observed differences in variability between BLR and continuum luminosities may be attributed to non-isotropic emission patterns. Although the total line luminosity does not change when the disk precesses, each segment of the BLR contributes to different segments of the broadened emission line (i.e., the redshifted or blueshifted side of the BLR may be more or less illuminated). We investigate these time-dependent asymmetries in the broad emission lines in Fig.~\ref{fig:halpha_spacetime}. In most panels shown, we depict contours of the $H\alpha$ flux in the $\delta v-t$ plane, where $\delta v$ is the frequency offset of H$\alpha$ converted to a velocity via the Doppler shift. In Fig.~\ref{fig:halpha_spacetime}a, we show the total strength of the line, which tracks the H$\alpha$ line luminosity depicted in Fig.~\ref{fig:lightcurve}c. It is clear that the line flux is to some degree asymmetric. These asymmetries are far more apparent in the flux residuals, which we show in Fig.~\ref{fig:lightcurve}b. Here, we compute our residuals by subtracting from an average line profile with a time-dependent normalization,
\begin{equation}
    F_{H\alpha}(\delta v, t)\propto L_{H\alpha}(t) f_{H\alpha}(\delta v),
\end{equation}
where $L_{H\alpha}$ is the total luminosity of the H$\alpha$ emission lines \citep[this is similar to the analysis by][]{horne_2021}. We can see clear positive (green) and negative (blue) excesses that evolve in tandem with the disk. In Fig.~\ref{fig:halpha_spacetime}d, we highlight the residuals during the tearing event highlighted in Fig.~\ref{fig:raptor}, where each panel in Fig.~\ref{fig:raptor} is labeled here by horizontal pink lines. During this period, we can see that the excess flux starts on the blueshifted side but migrates to the redshifted side. This is because as the disk precesses, it illuminates one side of the BLR more than the other. This is extremely interesting, as it suggests that evolution in the red-to-blue-asymmetries of the broad emission lines during a changing-look event may be a smoking gun for disk tearing.


\section{Discussion}
\label{sec:disc}
\subsection{Observational Prospects}
\label{sec:disc:obs}

It is likely that multiple physical mechanisms power CSAGN. How can we determine that disk tearing is responsible? Our results suggest two main prospects, 
\begin{itemize}
    \item If a CSAGN is powered by disk tearing, there will be continuum variability on different timescales (listed here for a $10^8\,M_\odot\,{\rm BH}$). Firstly, tearing cycles empty and refill the inner sub-disk on a months-to-years long timescale, which drives the change of state. Secondly, when the disk tears, the inner sub-disk precesses on a timescale of weeks (Fig.~\ref{fig:raptor}), further modulating the luminosity along the line of sight \cite[We also refer the reader][ who also discussed these considerations.]{raj_2021}. Thirdly, we detected a QPO driven by radial oscillations of the inner disk on intraday timescales (Fig.~\ref{fig:qpo}). These signals are more or less apparent depending on the individual tearing cycle but, if detected, would amount to strong evidence that the disk is torn. We emphasize that these timescales likely scale linearly with the BH mass and also have a still unknown dependence on parameters such as the spin and tilt angle. 
    
    \item When the inner sub-disk precesses, it drives time-dependent, asymmetric illumination of the BLR. This drives time-evolving red-to-blue asymmetries (Fig.~\ref{fig:halpha_spacetime}). Detecting time-evolving red-to-blue asymmetries during a changing-look event, especially correlated with the continuum variability, would be strong evidence for disk tearing. Interestingly, time-evolving red-to-blue asymmetries, indicative of precession, have already been detected in a luminous, variable AGN \citep{horne_2021}. 
\end{itemize}
Disk tearing would be primarily associated with repeating CSAGN. So, if disk tearing signatures are detected in an AGN, observational campaigns should continue to monitor the source. Also, as mentioned, all the timescales we have discussed depend linearly on the BH mass, which should be taken under consideration when interpreting AGN variability.

The long-term evolution of strongly warped AGN disks is unknown. It is possible that disk tearing is transient, occurring for several cycles before settling to a steady-state configuration. Even if tearing is persistent, the warped disk will gravitationally torque the SMBH into alignment on timescales of $10^6-10^9\,{\rm years}$ \citep[the large range in timescales is due to the dependence on BH mass and Eddington ratio; see][]{perego_2009}. This suggests that tearing-induced CSAGN may only occur when the AGN is young. 

We have reported an accretion disk simulation with toroidal magnetic fields and, thus, no relativistic jets. However, disks with poloidal magnetic fields may also tear  \citep{liska_2021},  launching jets that evolve and precess in tandem with the disk. In systems such as these, tearing cycles may also be accompanied by evolution of the corona and/or radio jet \citep[e.g.,][]{ricci_2020}. This suggests that there may be a rich phenomonelogy connecting disk tearing with evolving hard X-ray and radio emission.

\subsection{Caveats}
\label{sec:disc:caveats}

We have made several assumptions in this work. Firstly, the GRMHD simulation we used is not explicitly radiative, and thus the thermodynamic state of the disk is not self-consistent. This affects both the scale height of the disk  and the temperature distribution, which determines the flux (see Appendix \ref{app:temp} for details). Given these uncertainties, we chose only to model optically thick, blackbody emission, and make no predictions about the coronal emission. Thus, there may be optically thin emission signatures that we are currently missing, such as from the hot tenuous gas that is launched when the disk tears. We have also parameterized the non-axisymmetric features in the disk (e.g., emission directly from the nozzles) in a simplified manner (see App.~\ref{app:temp} -- this choice did not significantly affect our results). However, nozzle shocks may lead to hard X-ray emission that we are unable to reliably predict without explicitly evolving radiation. We also performed our ray-tracing calculation in the fast light approximation, whereas it would be more self-consistent to ray-trace in the slow-light approximation (i.e., allowing the disk to evolve while the rays propagate). 

The structure of the BLR is also simplified and we have not performed a parameter space survey of BLR properties. Specifically, we fixed the number and column densities of the gas and assumed a torus-like geometry \citep[e.g., ][]{GRAVITY_2018}, whereas the BLR may sometimes be better described by a wind-like geometry \citep[e.g.,][]{murray_1995}. Additionally, we have neglected
thermal and turbulent motions of the BLR. These choices may change the resulting line luminosities. Still, despite these uncertainties, we expect our results to be qualitatively robust, as it is difficult to avoid large swings in both the continuum and broad line emission when the disk tears. 

\subsection{Summary}
\label{sec:disc:summary}
We have presented synthetic observations of an extremely high resolution  \hammer{} simulation of an accretion disk tilted by $65^\circ$ with respect to a rapidly rotating $10^8\,M_\odot$ SMBH over a nearly two year epoch. We created our mock observations by performing ray-tracing calculations with the \RAPTOR{} code. We used both a line of sight camera and an azimuthal distribution of cameras in a presumed torus-like BLR. After performing ray-tracing calculations at each BLR camera, we performed \CLOUDY{} calculations to determine the resulting photoionization state of the BLR gas, allowing us to predict line luminosities. These lines are broadened by the Keplerian rotation of the BLR gas. Our main findings are,
\begin{itemize}
    \item \textbf{Our simulated accretion disk underwent several tearing cycles, causing the luminosity to vary by over an order of magnitude on months-to-years long timescales, consistent with CSAGN.} During an event, the inner ($\lesssim10-20\,r_{\rm g}$) disk tears off from the outer disk and is then quickly accreted by the BH. When the tear happens, the luminosity is high, but soon afterwards the luminosity is low because the inner region is depleted of gas. Sometimes, the inner sub-disk transiently aligns with the BH, resulting in a lower luminosity. The tearing cycles do not repeat with a coherent periodicity.
    
    \item \textbf{The precessing inner sub-disk modulates the luminosity along the line of sight.} This modulation imprints a shorter timescale (weeks-long, in our case) during the peak of the emission \citep[see also][]{raj_2021}. Since the tearing event is violent, the inner disk survives for only a few periods. This may be a detectable signature of disk tearing. 
    \item \textbf{We detected a coherent, intraday QPO during one of our tearing events.} We attribute the QPO to radial epicyclic oscillations of the inner sub-disk \citep[e.g.,][]{gibwa_2023}. The oscillating gas is located near the tear at $r_0\sim14\,r_{\rm g}$ and is most visible at cooler temperatures, since hotter temperatures track smaller radii. 
    \item \textbf{The tearing cycles drive the disappearance and reappearance of broad emission lines.} This is apparent in the optical spectra, the integrated line luminosities, and the equivalent widths. This is fundamentally driven by the time-dependence of the ionizing flux from the inner disk. 

    \item \textbf{When the torn inner disk precesses, it drives time-evolving, red-to-blue asymmetries of the broad emission lines.} This is because the precessing disk alternates between illuminating the side of the BLR moving toward and away from the line of sight. Detecting this signal in nature would be extremely suggestive of a warped and torn accretion disk.
    \end{itemize}

\begin{acknowledgments}
We thank Jonatan Jacquemin and Sasha Tchekhovskoy for insightful discussions. We thank Yossi Shvartzvald for providing the \textit{ULTRASAT} filters. We thank Elias Kammoun for recommending the calculation of the equivalent widths. NK is supported by a joint PCTS/PGI postdoctoral fellowship. ML was supported by the John Harvard, ITC and NASA Hubble Fellowship Program fellowships, NASA ATP award 80NSSC22K0817 and NSF AAG award AST-2407809. 
An award of computer time was provided by the Innovative and Novel Computational Impact on Theory and Experiment (INCITE), OLCF Director's Discretionary Allocation, and ASCR Leadership Computing Challenge (ALCC) programs under awards PHY129 and AST178. This research used resources of the Oak Ridge Leadership Computing Facility, which is a DOE Office of Science User Facility supported under Contract DE-AC05-00OR22725.
\end{acknowledgments}

\appendix
\section{TEMPERATURE PRESCRIPTION}
\label{app:temp}
\begin{figure}
    \centering
    \includegraphics[width=\textwidth]{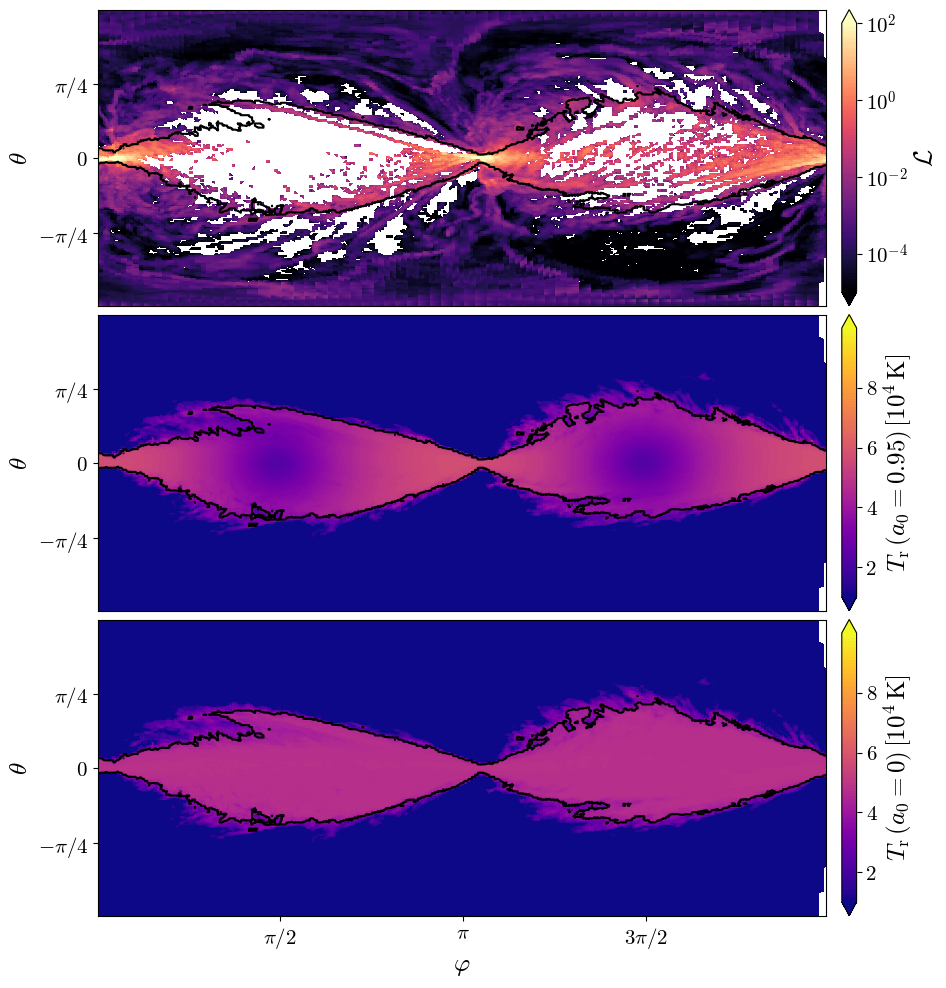}
    \caption{Thermodynamic properties of disk at $r=25\,r_{\rm g}$ in $\varphi-\theta$ plane, tilted such that the local disk midplane coincides with $\theta=\pi/2$ (see App.~\ref{app:temp} for details). \textbf{Panel a.} We show the cooling function used in the simulation (Eq.~\ref{eq:app:cool}). \textbf{Panel b.} We show our prescribed radiation temperature (Eq.~\ref{eq:app:tr}). \textbf{Panel c.} We show the radiation temperature without prescribed warp-dependent non-axisymmetries (Eq.~\ref{eq:app:tr} with $f_\psi=1$.}
    \label{fig:app:temperature}
\end{figure}

 We use the radiative transfer software RAPTOR \citep{RAPTOR} to ray-trace our H-AMR simulation. The equations of non-radiative ideal GRMHD are scale-free, so we must choose a BH mass and a density scale. We set the mass to $M=10^8\,M_\odot$ and set the density scale such that the accretion rate is near $\sim1\%$ of the Eddington limit (assuming a $10\%$ radiative efficiency). Before ray-tracing our results, we need to prescribe the emission and opacity coefficients to our simulation data. We assume that the disk is dominated by a scattering opacity and emits thermally. Since our simulation does not explicitly evolve radiation, we must prescribe the radiation temperature. We want to do this such that the resulting luminosity is consistent with our cooling function. We cool the disk to a target scale height on an orbital timescale using the function \citep{Noble_Cooling},
\begin{equation}
    \mathcal{L} = \Omega u_{\rm g} \sqrt{Y-1+|Y-1|},
    \label{eq:app:cool}
\end{equation}
Here, $\Omega=\sqrt{GM}{r_{\rm g}^3}\left((r/r_{\rm g})^{3/2}+a\right)$ is the frequency of circular orbits and $u_{\rm g}$ is the internal energy density. This function is nonzero when $Y>1$, where $Y=2(\gamma-1)u_{\rm g}/\pi H_{\rm th}^2\Omega^2\rho$ is the ratio of the internal energy to the target thermal energy which is set by $H_{\rm th}=0.02r$. In practice, the cooling function activates primarily at the nozzles (Fig.~\ref{fig:app:temperature}). We calculate the total flux leaving the disk at each radius,
\begin{equation}
\mathcal{F}(r)=\int \mathcal{L} dA_{\theta\varphi}/2\pi r
\end{equation}
We can use the flux to calculate the effective temperature of the disk at each radius using the Stefan-Boltzmann law,
\begin{equation}
    T_{\rm eff}= (\mathcal{F}/\sigma_{\rm SB})^{1/4} 
\end{equation}
If we set the radiation temperature to $T_{\rm eff}$ everywhere in the disk, \RAPTOR{} will produce a total luminosity consistent with our cooling function. However, we will have washed out non-axisymmetric emission (Fig.~\ref{fig:3d}). Since we cannot reliably estimate how non-axisymmetric the emission is (which depends, e.g., on shock heating in the nozzles), we modify the effective temperature using the function
\begin{equation}
    f_\psi = \left[(1-a)+2a(1-(\hat n\cdot\hat r)^2)\right]^{1/4},
\end{equation}
where we have used the unit vector $\hat n = \hat l\times r\partial_r\hat l$, where $\hat l(r)$ is the local angular momentum unit vector in the disk. The nozzles exist where $\hat n\cdot\hat r=0$, at which $f_\psi=1+a$. The amount of asymmetry is given by, 
\begin{equation}
    a={\rm min}((\psi/0.1)^2,a_0)
\end{equation}
where $\psi=|r\partial_r\hat l|$ is the warp amplitude and we have adopted the scaling $a\propto \psi^2$ since it is roughly consistent with how the thermal energy varies with azimuth in a warped disk \citep{kaaz_2025}. We divide $\psi$ by $0.1$ since this is of order the characteristic $\psi$ where bouncing sets in. In practice, we set the maximum asymmetry as $a_0=0.95$, but ultimately we found that our choice of $f_\psi$ did not strongly affect our results. Our radiation temperature is then, 
\begin{equation}
T_{\rm r} = T_{\rm eff}(r)f_\psi \exp(-\lambda_{\rm mfp}/4r),
\label{eq:app:tr}
\end{equation}
where we have also exponentially damped the radiation temperature in optically thin regions. Here, $\lambda_{\rm mfp}=1/\sigma_{\rm T} n$ is the local mean free path and $n$ is the number density. We do this because optically thin regions are unlikely to be thermal and we cannot reliably prescribe their thermodynamics. 

In Figure \ref{fig:app:temperature}, we sketch the thermodynamic properties of the disk at a fixed radius $r=25\,r_{\rm g}$ in a $\varphi-\theta$ frame that is rotated such that the local angular momentum vector of the disk is oriented vertically \citet[see][for details]{Kaaz_2023}. In Fig.~\ref{fig:app:temperature}a, we plot the cooling function Eq.~\ref{eq:app:cool}, which we can see peaks strongly at the nozzles. In Fig.~\ref{fig:app:temperature}b, we plot $T_{\rm r}$ assuming maximal asymmetry $a_0=0.95$ (which we use in the main results). By comparing panels a and b, we see that even at $a_0=0.95$, our cooling function is more strongly asymmetric than our prescribed radiation temperature, owing to shock cooling. This may imprint further substructure into, for example, the broad lines, but accurately predicting the radiation temperature at the nozzle shocks requires explicitly radiation transport. In Fig.~\ref{fig:app:temperature}c, we show the $a_0=0$ radiation temperature as a reference. In all three panels, the black curve is an estimate of the photosphere, defined by $\lambda_{\rm mfp}=r$. 

\bibliographystyle{aasjournal}
\bibliography{references}

\end{document}